\documentclass[prb,twocolumn]{revtex4}
\usepackage{graphicx}
\usepackage{latexsym}

\begin{document}

\title{
Vortex flow generated by a magnetic stirrer
}

\author{G\'abor Hal\'asz$^1$}
\author{Bal\'azs Gy\"ure$^1$}
\author{Imre M. J\'anosi$^1$}
\author{K. G\'abor Szab\'o$^{2,3}$}
\author{Tam\'as T\'el$^1$}

\affiliation{$^1$von K\'arm\'an Laboratory for Environmental Flows,
E\"otv\"os University, 
P\'azm\'any P. s. 1$/$A, H-1117 Budapest, Hungary}

\affiliation{$^2$Department of Fluid Mechanics, Budapest University of Technology and Economics, 
H-1111, Bertalan L. u. 4--6, Budapest, Hungary}
\affiliation{
$^3$Previous address: HAS Research Group for Theoretical Physics, 
H-1518, P.O. Box 32, Budapest, Hungary}%

\date{\today}

\begin{abstract}
We investigate the flow generated by a magnetic stirrer
in cylindrical containers  by 
optical observations, PIV measurements and particle and dye tracking methods.
The tangential flow is that of an ideal vortex outside of a core,
but inside downwelling occurs with a strong jet in the very middle.
In the core region dye patterns remain visible over minutes indicating 
inefficient mixing in this region.    
The results of quantitative measurements
can be described by simple formulas in the investigated region
of the stirring bar's rotation frequency. The tangential flow 
turns out to be dynamically similar to that of big atmospheric vortices 
like dust devils and tornadoes.
\end{abstract}

\maketitle

\section{Introduction}

Magnetic stirrers are common 
equipments in 
several types 
of laboratories. The main component of this equipment is a 
magnet rotating with adjustable frequency around a fixed 
vertical axis below a flat horizontal surface.
The rotation of this magnet brings a magnetic stirrer bar in rotation
on the bottom of a container put on the flat surface.   
If the container is filled with a liquid, the bar 
generates strong fluid motion which is believed to cause
efficient stirring and mixing. A striking pattern of such flows 
is a  
big vortex above the stirrer bar and the 
corresponding depletion, the funnel, on the surface,
which indicates that the flow is strongly structured. 

Our aim is the experimental investigation of the 
fluid motion in cylindrical containers generated by  magnetic stirrers.
Surprisingly enough, despite the widespread use of 
this device we could not find any reasonable 
description of such flows in the literature.

In the next Section simple theoretical models of isolated vortices 
are reviewed.
Then (in Sections III, IV) we describe the experimental setup and 
the used methods of data
acquisition.  In Section V we present
the results of optical observations, of PIV measurements and of
monitoring tracer particles and dyes.
Section VI is devoted to deriving simple relations 
for the vortex parameters based on the measured data. 
The concluding 
section points out the similarities and the differences of our results 
compared with other whirling systems: bathtub vortices, dust devils 
and tornadoes.

\vspace*{-0.5cm}
\section{Theoretical background}
\label{2.3}

Here we review the most important elementary models of steady isolated vortices
in three-dimensional fluids of infinite extent.\cite{LL,Lugt,Laut}
Although our system 
is obviously
more complicated then these models, they are useful reference points in
interpreting the data. The model flows shall be expressed in
cylindrical (radial, tangential and axial) velocity components ($v_r$, $v_t$,
and $v_z$).

{\em Rankine vortex.}
In this model the vorticity is uniformly distributed in
a cylinder of radius $c$ with a central line (the $z$ axis) as its axis.
The tangential component is continuous in $r$ but a break appears
at the radius $c$. The two other components remain zero:
\begin{equation}
v_r=0, \;\; v_t=\frac{Cr}{c^2}, \mbox{for} \;\; r\le c, \;\;
v_t=\frac{C}{r}, \mbox{for} \;\; r>c, \;\; v_z=0.
\end{equation}
Within the radius $c$ a rigid body rotation takes place, while outside
a typical $1/r$-dependence appears with $C$ proportional to the
circulation of the flow. 
The limit $c \rightarrow 0$ corresponds to 
an ideal vortex line.

{\em Burgers vortex}.
In a real fluid viscosity smoothes out the
break in the tangential component of the Rankine vortex.
In order to maintain a steady rotation, an inflow and an
axial flow should be present. In the Burgers vortex the strength
of the axial flow increases linearly with the height $z$, measured from a
certain level:
\begin{equation}
v_r=-\frac{2 \nu}{c^2}r, \;\;\;\;
v_t=\frac{C}{r}\left( 1-e^{-r^2/c^2} \right), \;\;\;\; v_z=\frac{4 \nu}{c^2}z.
\label{Burg}
\end{equation}
Here $\nu$ is the kinematic viscosity of the fluid, and $c$ remains
an effective
radius within which the tangential flow is approximately a
rigid body rotation.
Note that the tangential velocity component may depend on the viscosity 
indirectly only, via
a possible $\nu$-dependence of the radius $c$.

\section{Experimental setup}

\begin{figure}
\resizebox{0.25\textwidth}{!}{\includegraphics{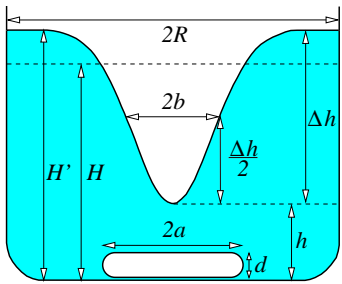}}
  \caption{
The most important optically observable  
parameters of the problem. Geometrical ones:  $R$, the radius of the
cylinder, $a$ and $d$ parameters of the stirrer,
and $H$, the height of the still water.
While stirring, further observable parameters are
the displaced water 
height
$H'$, the distance $h$ between the funnel's
deepest point and the bottom of the container,
the funnel's depth and halfwidth $\Delta h=H'-h$ and $b$, respectively.
}
  \label{F:1}
\end{figure}

The experiments were carried out with tap water in glass cylinders,
in different combinations 
of radii $R$ and initial water heights $H$ 
as summarized in Table \ref{T:1}.

\begin{table}
\centering
\begin{tabular}{|c|c|}
  \hline
  $R$ (cm) & $H$ (cm) \\  \hline
  $3.8$ & $12.0$ \\
  $6.5$ & $16.8$ \\
  $10.5$& $24.8$ \\
  $22.4$& $27.1$ \\ \hline 
\end{tabular}
\caption{The radii $R$ of the cylinders used and the
heights $H$ to which water was filled up in each.} 
\label{T:1}
\end{table}

Two other relevant geometrical parameters are provided by the  
length  $2a$ and
the width $d$ of the magnetic stirrer bars
(see Fig. \ref{F:1}). The different parameters of the
bars used are summarized in Table \ref{T:2}.     

When the rotation of the magnetic stirrer is switched on, the water column starts
moving and after some time (which is on the order of minutes in our case)
a statistically stationary flow sets in. 
Inherent fluctuations around the mean
arise due to the periodic motion of
the stirrer bar, the lack
of a fixed axis of rotation,
and turbulence. These are the physical reasons behind
the relative errors of our measurements being on the order of $10$ percents.
The most striking optically observable object is the funnel developing on the
water surface.  The height $H'$ of the free surface at the perimeter and
the characteristic sizes of the funnel,
as defined in Fig. \ref{F:1}, can be easily measured
(see Section \ref{4.4}).

\section{Data acquisition}
\label{4.4}

\begin{figure}
\resizebox{0.48\textwidth}{!}{\includegraphics{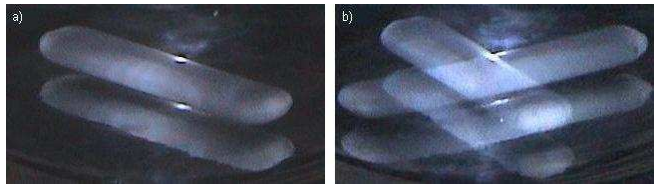}}
  \caption{
The apparent picture of the bar (along with its mirrored
image by the bottom) at stroboscopic frequencies
$f_1$ (a) and $f_{1/2}$ (b).
}
  \label{F:2}
\end{figure}

{\em The rotation frequency} $\Omega$ of the stirring bar in the container filled up 
with water
was determined by means of a
stroboscope whose frequency $f$ is adjustable in a broad range. 
At certain frequencies $f_n$ the bar appears to be at rest 
(see Fig. \ref{F:2}a).
This happens if the bar rotates an integer multiple
of a half rotation between two flashes, i.e., if  
\begin{equation}
\frac{\Omega}{f_n} =n \pi.
\end{equation}
The largest value of the $f_n$-s uniquely determines the bar's rotation
frequency as
$\Omega=f_1 \pi$. In order to reach a higher accuracy, we also
determined the frequency $f_{1/2}$ when   
the bar rotates a right angle between two flashes. This case is
designated by the
appearance of a steady cross traced out by the bar on the bottom of the container
(see Fig. \ref{F:2}b). The bar's rotation frequency was determined as the average
of the frequencies belonging to $f_1$ and $f_{1/2}$. The uncertainty in $f$
is $0.1$ Hz. In the range $(20, 120)$ s$^{-1}$ of $\Omega$ investigated, this 
corresponds to a relative 
error of about $1$ percent, which is negligible compared to the other errors.   

\begin{table}
\centering
\begin{tabular}{|c|c|c|c|}
  \hline
stirrer bar &  $a$ (cm) & $d$ (cm) &symbol \\  \hline
i &  $2.05$& $0.85$ & $\bigtriangledown$\\
ii&  $2.50$& $0.90$ & $\Box$ \\
iii &  $4.00$& $1.00$ & $\Diamond$ \\ \hline
\end{tabular}
\caption{Geometrical parameters of the stirrer bars and the
symbols used to mark the corresponding measured data in Figs.
\ref{F:5} and \ref{F:13}.
}
\label{T:2}
\end{table}

{\em The funnel parameters}  were determined via direct optical observations.
The water height $H'$ in the statistically
stationary
 state was measured by means of a 
ruler. The height of the funnel's deepest point can be determined in a 
similar way or
by using a horizontal laser sheet. From these two quantities, the funnel depth
is simply $\Delta h=H'-h$ (cf. Fig. \ref{F:1}).    
The halfwidth of the funnel was measured on the back side of the cylinder. 
The length
obtained this way 
was corrected by taking into account the optical effect caused by the
cylindrical lens formed by the water column to obtain the value $b$. 
Both the funnel depth and width are subjected to an error of about
$10$ percents due
to the fluctuations, c.f., Section III.

{\em The particle image velocimetry} (PIV) method was used to determine the velocity 
field in a plane defined by 
a horizontal 
laser sheet. From the position of
fine tracer particles on two subsequent images taken with a time difference
of about $10^{-2}$ s, the displacement and velocity of the particles can be 
determined.\cite{Dant} We used a commercial PIV equipment (ILA GmbH, Germany) 
and determined the
flow field in the largest container ($R=22.4$ cm) at an intermediate
water height ($H=16.8$ cm) with  stirrer bar ii (cf. Table \ref{T:2})
in different horizontal layers.     
The presence of the funnel and the strong downdraft make the PIV data 
unreliable in the middle of the container, within a region of radius of about 
$8$ cm.  

{\em Particle tracking} enables us to
study the central region of the flow. 
We used plastic beads (low density polyethylene) 
of diameter $\sim 1$ mm,
which has a density of $\sim 0.92$ g/cm$^3$. 
Despite being lighter than water,
they sink below
the funnel and often reach a 
dynamical steady state (for more detail see 
Subsection V.C). The approximate strength of the downwelling
in the middle
was estimated as the rising velocity of the beads
in a water column at rest. From several measurements
in a separate narrow glass cylinder we found this rising
velocity to be $7$--$8$ cm/s, for all the beads used.  

{\em Spreading of dye}
can provide
a qualitative picture about the flow.  
A particularly important region 
is that around the axis of the vortex, where this coloring technique reveals 
fine details (see Subsection V.D).

\section{Results}

\begin{figure}
\resizebox{0.45\textwidth}{!}{\includegraphics{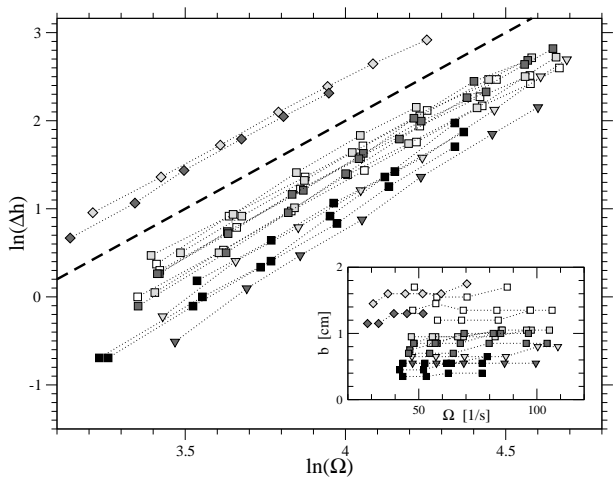}}
  \caption{
Funnel depth as a function of the rotation frequency 
in containers of different 
radii for different water heights and stirrer bars
in double logarithmic representation. The slope of the dashed line
is $2$. 
Inset: Funnel halfwidth as a function of the
rotation frequency in different measurements. 
Different symbols mark different stirring bars
(cf. Table \ref{T:2}), 
and 
the level of grayness increases with
the radius of the container.
Symbols are not distinguished according 
to water heights.
}
  \label{F:5}
\end{figure}

{\bf A. Frequency dependence of the funnel.}
The results  of the optical observations of $16$ cases (different containers, water heights and 
stirring bars) each measured at several frequencies $\Omega$ are summarized 
in Fig. \ref{F:5}. While there is a clear frequency dependence
on the funnel depth $\Delta h$, the halfwidth $b$ appears to be independent
of $\Omega$ (see inset), in first approximation at least.  

To extract the form of the frequency dependence, the funnel depth is
plotted on log-log scale in Fig.\ \ref{F:5}. The straight lines
clearly indicate a power-law dependence. The exponent is read off 
to be $2$:
\begin{equation}
\Delta h \sim \Omega^2.
\label{hO2}
\end{equation}   
One sees that the coefficient (not written out) depends much stronger on 
the stirring bar's parameters $(a,d)$  
than on the container's geometry ($R,H$).

\begin{figure}
\resizebox{0.45\textwidth}{!}{\includegraphics{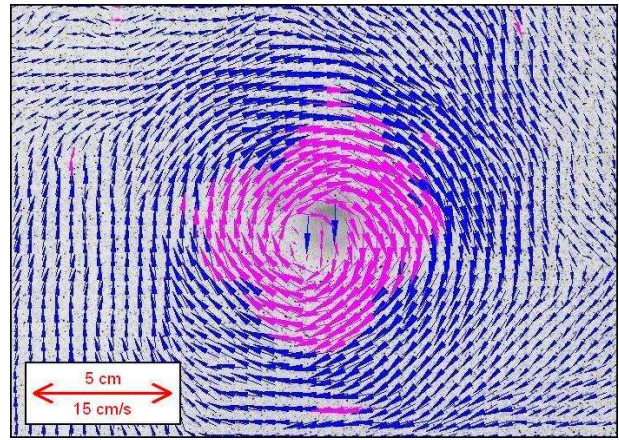}}
  \caption{
Detail of a PIV image at the height of $z=13$ cm taken at rotation
frequency $\Omega=35.5$ s$^{-1}$ ($R=22.4$ cm,
$H=16.8$ cm, $a=2.5$ cm, $d=0.9$ cm).  
The inset sets the length and velocity scales. 
}
  \label{F:6}
\end{figure}

{\bf B. Velocity fields.}
First we present, in Fig. \ref{F:6}, the result of a typical PIV measurement
in a horizontal plane. The arrows mark velocity vectors. Arrows 
around the vortex center 
are produced by algorithmic interpolation and cannot be 
considered therefore to be quantitatively
reliable. The flow is not fully axially symmetric,
small secondary vortices appear around the edges of the picture.

\begin{figure}
\resizebox{0.48\textwidth}{!}{\includegraphics{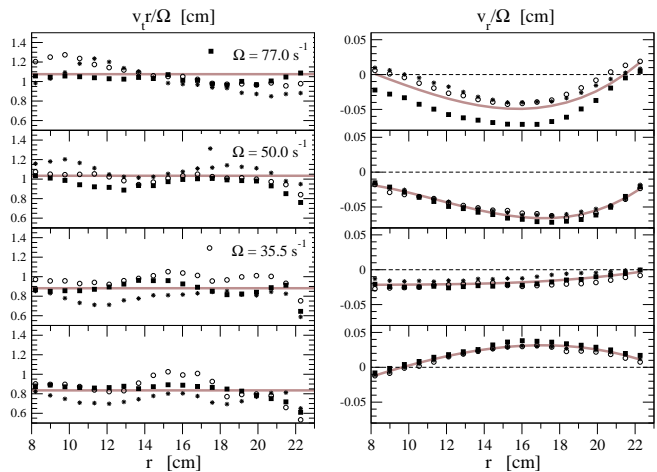}}
  \caption{
Rescaled tangential (left column) and radial (right column)
velocity components $v_tr/\Omega$ and
$v_r/\Omega$ for three values of $\Omega$ (see
legends).
PIV measurements carried out at different heights
$z=$ 16, 13, 8, and 4 cm,
from top to bottom. The horizontal lines on the left are fitted average values.
The smooth curves on the right
join the neighboring measured
points to guide the eye.
}
  \label{F:7}
\end{figure}

In order to understand the mean flow, we divided each PIV image into 
narrow concentric rings and 
evaluated the average tangential and radial velocity 
in each band. These values were further averaged over several images taken 
at different times in the same flow and at the same height. This way the effect
of secondary 
vortices visible in Fig. \ref{F:6} became averaged out. 
The procedure leads to a
discrete representation of the functions $v_t(r)$ and $v_r(r)$. 
Both components appeared to be proportional to the frequency of the stirring 
bar, therefore we 
present in Fig. \ref{F:7} 
the components already divided by 
$\Omega$. Since the results in the innermost region are not reliable, 
the components are displayed for distances $r>8$ cm only. 

The measured tangential velocity data $v_t(r)$
can be approximated by a functional
form $A/r$ satisfactorily. This behavior is
demonstrated by plotting the
rescaled quantity $v_t r/\Omega$
in Fig. \ref{F:7}, left column. (Note that
the multiplication by $r$ magnifies the
apparent error.)
The coefficient $A$ is a weakly 
decreasing function of the depth. 
Nevertheless, the
tangential flow matches that of an ideal vortex,
as a first approximation. Accordingly, the vortex strength $C$ is
proportional to the frequency, i.e.,  
\begin{equation}
C=A \Omega
\label{CAO}
\end{equation}
with a coefficient $A$ extracted from the data to be $A=0.9\pm 0.2$ cm$^2$. 

In contrast to the tangential component, the radial component depends 
strongly on the height (Fig. \ref{F:7},
right column). In the upper layers there 
is an inflow ($v_r<0$) for
$r>8$ cm at least, which decays towards zero as the height decreases. The level
$z=4$ cm is dominated by outflow, but around $r=8$ cm a weak inflow survives. 

The planar PIV algorithm does not provide any information on the vertical 
velocities. This component can, however, be determined from the continuity 
equation which takes the form \cite{LL,Laut}
\begin{equation}
\frac{\partial v_z}{\partial z}=-\frac{1}{r} \frac{\partial ( r v_r)}{\partial r}
\end{equation}
in axisymmetric flows. We numerically integrated the left hand side
to get an approximation for $v_z(r)$ from the radial component. 
This component turns out 
to be proportional to $\Omega$, too. 
The qualitative feature is that there is upwelling in the outermost
$4-5$ cm and a slow downwelling in the intermediate region. 
The strong upwelling around the boundary of the
container implies that there must be a strong downwelling in the very 
center, which, however, cannot be resolved by means of the PIV method.

\begin{figure}
\resizebox{0.45\textwidth}{!}{\includegraphics{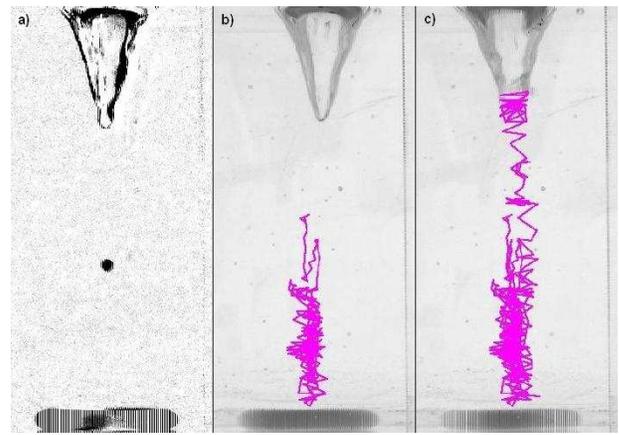}}
  \caption{
Particle tracking. a) A plastic bead lighter than water becomes
washed downwards below the funnel. b) Path of the particle over $10$ s.
c) Path of the particle over $20$ s. 
}
  \label{F:10}
\end{figure}

{\bf C. Particle tracking.}
The strength of downwelling in the very center of the vortex can be 
estimated by means of monitoring plastic beads, lighter than water. 
They float on the surface but become eventually trapped by the funnel, on 
the surface of which they slide down and become advected downwards toward
the bulk of the fluid (Fig. \ref{F:10}a).
Along the axis of the vortex the particles reach a 
dynamical steady state (subjected, of course, to considerable fluctuations). 
This
indicates that there is a strong downward jet in which the downward
drag acting on the 
particle approximately compensates the upward resultant of gravity and
buoyancy. 
As mentioned in Section
\ref{4.4}, the asymptotic rising velocity of the beads is $7-8$ cm$/$s 
in a water of
rest. Therefore we conclude that the strength of the downward jet is also 
$7-8$ cm$/$s.

A necessary condition for vertical stability is that the
velocity of downwelling decreases when moving downwards along the axis. This occurs unavoidable
in our case since the velocity should approach zero close to the lower bottom of the
container. Experience shows that the 
average vertical position is shifted downwards when 
the frequency $\Omega$ is increased, since this makes the jet somewhat stronger.

Stability is maintained in the horizontal direction as well.  This is due 
to the fact
that in a rotating system an `anticentrifugal' force acts on the bead
since it is lighter than the surrounding fluid. Whenever the particle 
deviates from the
axis, this force directs it backwards. This is accompanied by an immediate 
rising of the particle which indicates that the width of the strong 
downward jet is of the
same order as the diameter of the particles. We conclude that this
width is a few mm.    
 
Due to the presence of all these effects and the permanent fluctuations
of the flow, the overall motion of a particle is rather complex.
After leaving the central jet, it starts rising but the 'anticentrifugal' force
pushes it back towards the center, at a larger height. Then it is captured 
by the jet and starts moving downwards again, and will leave the jet 
at another
level than earlier. The bead remains within a cylindrical region around
the vortex axis.  
We conclude that the particle motion is chaotic \cite{Ott}
(Fig. \ref{F:10}b). By tracking a single particle over a long period of time,
a chaotic attractor is traced out
(Fig. \ref{F:10}c).

\begin{figure}
\resizebox{0.45\textwidth}{!}{\includegraphics{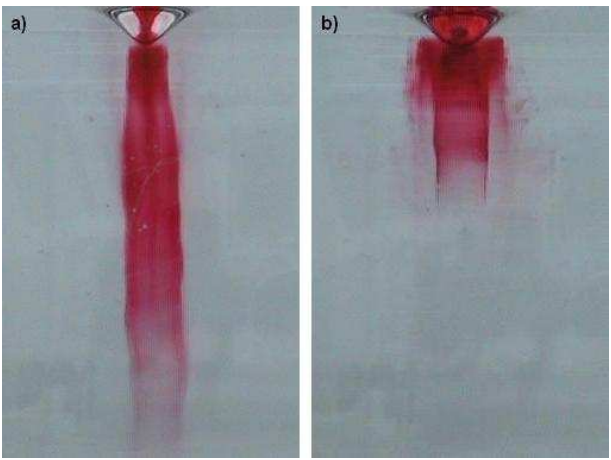}}
  \caption{
Dye pattern. (a) A stable cylindrical dye curtain
(of lifetime over a minute) develops 
below the funnel. (b) Often sublayers can also be observed.
}
  \label{F:11}
\end{figure}

{\bf D. Spreading of dye.}
When injecting dye into the water outside of the central region one 
observes a fast spreading. A drastically different behavior is
found in the middle of the funnel. Very quickly,
a cylindrical dye curtain develops around the vortex axis
which remains observable over about
a minute (Fig. \ref{F:11}). This can be seen at any 
value of the parameters ($R,H,a,d,\Omega$) investigated.
The radius of the dye cylinder is on the order of the magnitude of the
halfwidth $b$, it is about $1$ cm. It is remarkable that
the wall of the cylindrical region
containing the trapped beads (Fig.~\ref{F:10}c) 
coincides with this dye curtain.


The existence of this long-lived dye curtain 
indicates that the radial velocity is approximately
zero in a cylindrical annulus
around the axis of rotation, the diameter of which
is proportional to the width $d$ of the stirring
bar. Since, while rotating,
the bar continuously blocks the downward flow in a circle
of diameter $d$, the suction is the strongest
around the perimeter of this region.
The downwelling has a maximum strength here, and it
weakens inwards. In the very center, however, there is
the central jet mentioned in Subsection V.C,
therefore the existence of a local minimum is necessary.
The injected dye accumulates
along the surface in which the downward velocity takes 
its local minimum.
Thus both inside and
outside the cylinder surface the
downward flow is stronger than on the surface itself,
where injected dye accumulates.
When injecting the dye somewhat off the cylinder one often observes
more than one concentric
curtains, as well (Fig. \ref{F:11}b).
This indicates that
there might be more local minima of the downward velocity in a region
around the halfwidth of the funnel.

\begin{figure}
\resizebox{0.25\textwidth}{!}{\includegraphics{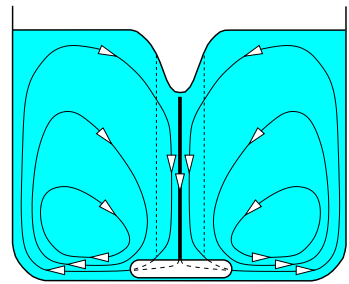}}
  \caption{
Pattern of the time averaged flow. The thin continuous lines
represent  streamlines, the bold line denotes the downward jet, and the
dashed line represent the average location of dye curtains.  
}
  \label{F:12}
\end{figure}

{\bf E. Qualitative picture.}
Based on the observations and the measurement of the velocity components,
we obtain the following qualitative 
picture of the time averaged flow (Fig. \ref{F:12}). In a given 
vertical slice two flow cells are formed by the upwelling at the 
outer walls and the downwelling in the middle. In the three-dimensional
space this corresponds to a torus flow whose central line lies in a  
horizontal plane, below the half of the water height. 
The flow along the central line has no radial and vertical 
component, a pure rotation takes place. 
The most striking feature is a strong downward jet in a very narrow
filament on the axis of the vortex
surrounded by a cylindrical region of
weak downwelling.

\section{Quantitative results}

{\bf A. Funnel depth.}
To obtain a simple expression for the funnel depth, we apply 
 the method of dimensional analysis \cite{K,X,Y} first, 
and a comparison with the measured
data leads then to a particular form.
As dimensionless measures of the viscous and 
gravitational 
effects in the rotating flow, a Reynolds and a Froude number is introduced
as
\begin{equation}
Re = \frac{\Omega a^2}{\nu}, \;\;\;\; \mbox{and} \;\;\;\;
Fr = \frac{\Omega a}{(g R)^{1/2}}\frac{d}{a},
\label{ReFr}  
\end{equation} 
respectively. (The usual Froude number is
multiplied by the factor $d/a$, in order
to simplify the argumentation below.)
Small values of them indicate strong viscous and gravitational effects.
With our typical data ($\Omega= 50$ s$^{-1}$, $R=10$ cm, $a=2.5$ cm, $d=1$ cm,
$\nu=10^{-2}$ cm$^2$s$^{-1}$) we obtain $Re=3 \cdot 10^4$ and $Fr=0.5$.
This indicates that gravity is essential, but viscosity is not so important
for the overall flow. It is, however, obviously important on small scales, like
e.g. in the center of the vortex.   
   
The ratio $\Delta h/d$ must be a function of the dimensionless parameters, 
therefore we can write
\begin{equation}
\frac{\Delta h}{d}= f\!\left( Re, Fr, \frac{a}{d}, \frac{a}{R}, \frac{H}{R} \right)
\end{equation} 
with $f$ as an unknown function at this point.
There might other dimensionless numbers also be present. One candidate would 
be a measure of the surface tension. 
This effect we estimated 
in control experiments with surfactants,
and found a negligible 
influence with respect to our typical
measurement error. Therefore, 
we do not include the corresponding dimensionless 
number into $f$. 

\begin{figure}
\resizebox{0.38\textwidth}{!}{\includegraphics{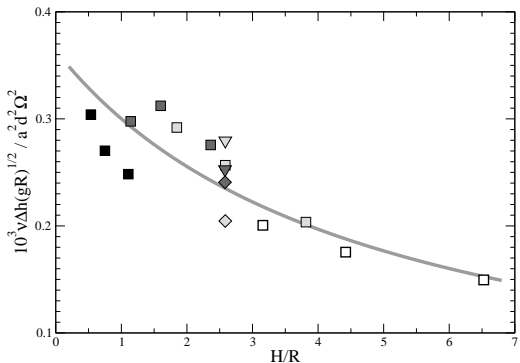}}
  \caption{ 
Determining the scaling function $\Phi(H/R)$ from the scaled measured 
$\Delta h/\Omega^2$ values 
averaged over different 
$\Omega$ values. The continuous line is a fitted hyperbola
of the form of (\ref{Phi}). The fact that several measured values belong
to a given $H/R$ is due to the use of different stirrer bars. 
The use of symbols is similar as in Fig. \ref{F:5}.
}
  \label{F:13}
\end{figure}

Assuming that $f$ is linear in both $Re$ and $Fr$ we obtain     
\begin{equation}
\frac{\Delta h}{d}= Re\, Fr \; \Phi\left(\frac{a}{d}, \frac{a}{R}, 
\frac{H}{R} \right). 
\end{equation} 
This assumption is supported not only by (\ref{hO2}), but by other observations
as well: a careful
investigation of the data shows that the funnel depth is proportional to $a^2$,
and for sufficiently large radii ($R > 0.4 H$) it scales as $R^{-1/2}$. 
These observations imply
that no $a/d$ and $a/R$-dependence remains in $\Phi$:       
\begin{equation}
\frac{\Delta h}{d}= 
\frac{\Omega^2 a^2 d}{\nu (g R)^{1/2}}\;
 \Phi\!\left( \frac{H}{R} \right).
\end{equation} 
The form of the single-variable function $\Phi(x)$ can be estimated from a 
replotting of the data, as shown in Fig. \ref{F:13}. As the fitted smooth
curve shows, a reasonable form of $\Phi$ is
\begin{equation}
 \Phi\left( x \right)= \frac{1}{\alpha x+k}.
\label{Phi}
\end{equation} 
The best choice of the parameters is $\alpha=(0.58 \pm 0.08)\cdot 10^{3}$,
 $k=(2.8 \pm 0.2) \cdot 10^3$.
The direct expression for the funnel depth is then
\begin{equation}
\Delta h= 
\frac{\Omega^2 a^2 d^2 R^{1/2}}{\nu (\alpha H+kR) g^{1/2}}.
\label{Dh}
\end{equation} 
The result shows that
the dependence on the water height is rather weak. This explains afterwards
why it was worth
defining the Froude number with $R$ in (\ref{ReFr}).

It is remarkable that such  a simple formula can be found to the
measured data with about $10$ percent accuracy.
 
{\bf B. Halfwidth.}
The halfwidth $b$ was found in Subsection V.A to be independent of $\Omega$.
The ratio $b/d$ can therefore be written as 
\begin{equation}
\frac{b}{d}= \frac{Fr}{Re} \, \Psi\!\left(\frac{a}{d}, \frac{a}{R}, \frac{H}{R} \right) =
\frac{d \nu}{a^2 (g R)^{1/2}}
 \Psi\!\left(\frac{a}{d}, \frac{a}{R}, \frac{H}{R} \right).
\end{equation} 
The data provide an essentially $H$-independent halfwidth which scales
with $R^{-1/2}$. Therefore $\Psi$ must be independent of $H/R$ and $a/R$,
i.e.,  
\begin{equation}
\frac{b}{d}= 
\frac{d \nu}{a^2 (g R)^{1/2}}
 \Psi\!\left(\frac{a}{d} \right).
\end{equation} 
Since the data show that $b$ is approximately 
linearly proportional to $a$,
function $\Psi(x)$ should be cubic:
\begin{equation}
 \Psi(x) = \beta x^3,
\end{equation} 
and the best fit yields $\beta=(2.8\pm 0.8) 10^3$. The direct expression
for the halfwidth is then
\begin{equation}
b= 
\frac{\beta a \nu}{d (g R)^{1/2}}.
\label{b}
\end{equation} 
Note that this expression does not contain the water height at all.

{\bf C. Interpretation in terms of a Burgers vortex.}
The fact that the rotating motion in the magnetic stirrer flow is accompanied
with an inflow and a downwelling resembles to the Burgers model
treated in Section \ref{2.3}. 
Equation (\ref{Burg}) defines the plane 
$z=0$ as a plane without vertical velocity.
Since far away from the center there is no downwelling on the free surface
at height $H'$, 
the $z=0$ level should be chosen as the topmost level 
of the rotated water. 
The vortex model obtained this way corresponds to the bulk
of the investigated flow, far away from the external walls and the stirrer bar.
It does not describe either the upwelling near the walls or the outdraft
around the stirring bar.

The time independent
nature of the axisymmetric mean flow implies that the pressure gradient
compensates the centrifugal force and gravity in the radial and vertical 
direction:
\begin{equation}
\frac{\partial p}{\partial r}= \varrho \frac{v_t^2}{r},\;\;\;\;
\frac{\partial p}{\partial z}= - \varrho g,
\end{equation}
where $\varrho$ is the fluid density. 
On the fluid's free surface at $z=\eta(r)<0$ $d \eta/d r =  
-({\partial p}/{\partial r})/
({\partial p}/{\partial z})$. Consequently,
\begin{equation}
\frac{d \eta}{d r} =
 \frac{v_t^2}{rg}.
\end{equation}
By inserting here the tangential velocity component from (\ref{Burg}), 
the funnel depth can be obtained by integration:
\begin{equation}
\Delta h  =\int_0^{R'} 
 \frac{C^2}{r^3 g}\left(1-e^{-r^2/c^2}\right)^2 dr.
\end{equation}
Here $R'\gg c$ is the radius within which the Burgers model is valid.
Due to the exponential cut-off within the integrand the integral can well be
approximated by taking $R'=\infty$.  We thus obtain
\begin{equation}
\Delta h  =\ln{2}  
 \frac{C^2}{c^2 g}.
\end{equation}

By equating this with the funnel depth expression (\ref{Dh}) derived above,
we recover relation $C=A \Omega$ found in Subsection V.B with a specific
coefficient
\begin{equation}
A = a^2 \left( \frac{ \beta^2 \nu}{\ln{2} (\alpha H+kR) (gR)^{1/2}}.
            \right)^{1/2}.
\label{A}
\end{equation}
Similarly, from $c \approx b$ and the halfwidth expression (\ref{b}) 
based on the measured data, we obtain 
\begin{equation}
c= 
\frac{\beta a \nu}{d (g R)^{1/2}},
\end{equation} 
a relation already used in (\ref{A}).
Thus we are able to express both basic parameters, $C$ and $c$ 
of the Burgers model
with the directly measurable parameters of the flow investigated. 

\vspace*{-0.5cm} 
\section{Discussion}

Here we compare the fluid dynamical properties of
our experiment with that of other whirling systems:
bathtub vortices, dust devils and tornadoes.

Detailed measurements of the velocities 
in these flows 
(see Refs. \onlinecite{ABSRL}, \onlinecite{Greeley}, 
and \onlinecite{SHGLW}-\onlinecite{SH},
respectively)
indicate that, in spite of basic differences in the 
other components, 
the tangential component outside of the vortex core 
decays with distance $r$ as $C/r$, independently of height.
The most dominant tangential component of all the flows is thus
practically identical.

To estimate the degree of dynamical similarity in this component, we use
another set of the dimensionless numbers:   
\begin{equation}
Re' = \frac{U c}{\nu}, \;\;\;\; Fr' = \frac{U}{(g H)^{1/2}}.
\label{Re1Fr1}
\end{equation} 
Here $U$ represents the maximum velocity of the 
tangential flow component and
$c$ denotes the radius of the vortex core.

\begin{table}
\centering
\begin{tabular}{|c|c|c|c|c|c|}
\hline
flow       &  $U$ (m$/$s)    & $H$ (m) & $c$ (m)           & $Re'$          & $Fr'$  \\  \hline
stirrer    &  $0.5$          & $0.2$   & $10^{-2}$         & $5 \cdot 10^3$ & $0.4$  \\
bathtub    &  $0.4$          & $0.1$   & $2 \cdot 10^{-4}$ & $8 \cdot 10^2$ & $0.4$  \\
dust devil &  $25$           & $10^3$  & $50$              & $8 \cdot 10^7$ & $0.25$ \\
tornado    &  $70$           & $10^3$  & $200$             & $10^9$         & $0.7$  \\ \hline
\end{tabular}
\caption{Parameters and dimensionless numbers for the tangential velocity components of the flows compared.}
\label{T:3}
\end{table}

The value of $U$ can be estimated
in our case to be a few dm$/$s, for the estimate we take  $50$ cm$/$s.
The core radius and the height are $c \approx b \approx 1$ cm and
$H \approx 20$ cm,
respectively. The data for the other flows are taken from the literature
and are
summarized, along with the resulting dimensionless numbers in Table \ref{T:3}.

The Froude numbers are on the same order of magnitude but the
Reynolds numbers are rather different. This shows that the role of 
viscosity is much stronger in small scale flows then in the atmosphere. 
Viscous flow
is present in the vortex core, therefore the detailed flow patterns 
do not match there. The global flow is, however, in all cases 
practically that of an ideal fluid.    
Therefore we conclude that outside of the vortex core the tangential flows
are dynamically similar, i.e. our experiment faithfully models
all these whirling systems, an observation which can be utilized 
in undergraduate teaching.

Finally we mention that the analog
of the dye curtain can be seen in tornadoes, typically in the vicinity
of the bottom since it is the Earth surface which
is the source of `dye' (in the form of dust or debris).
In some tornadoes this `dye' curtain is clearly separated
from the funnel
(cf. e.g., http:$//$www.oklahomalightning.com).

This work was supported by the Hungarian Science Foundation
(OTKA) under grants TS044839, T047233. IMJ thanks for a
J\'anos Bolyai research scholarship of the Hungarian
Academy of Sciences.

\section*{Appendix}

Student Project 1. Explore the advection properties
of small objects of different sizes, shapes and
densities. Observe the rotation of elongated bodies
(e.g., pieces of plastic straws) 
in the central jet. Their angular velocity monitors
the local vorticity of the flow.

Student Project 2. As mentioned,
one source of experimental errors is the oscillation
of the stirrer bar. 
Try to eliminate this effect by slightly modifying the setup.
Hint 1.
Insert two magnetic stirrer bars in the opposite ends
of a plastic tube with a hole between the two magnets.
Use an external frame to fix a thin rod led through
the hole to obtain a fixed axis of rotation.
Hint 2. 
Use a narrow cylinder of nearly the same
diameter as the length of the bar.
Estimate the magnitude of the error
and compare the parameters of the funnel
in the modified and original setups.

\end{document}